# Preprint Virtual Reality GIS and Cloud Service Based Traffic Analysis Platform

Xiaoming Li[1,2,3], Zhihan Lv[3], Weixi Wang[1,2], Chen Wu[4], Jinxing Hu[3]
1. Shenzhen Research Center of Digital City Engineering, Shenzhen, China
2. Key Laboratory of Urban Land Resources Monitoring and Simulation,
Ministry of Land and Resources, Shenzhen, China
3. Shenzhen Institute of Advanced Technology(SIAT), Chinese Academy of Science, Shenzhen, China
4. State Key Laboratory of Information Engineering in Surveying, Mapping and Remote Sensing,
Wuhan University, Wuhan, China
Email: lvzhihan@gmail.com

*Abstract*—City traffic data has several characteristics, such as large scale, diverse predictable and real-time, which falls in the range of definition of Big Data. This paper proposed a cloud service platform which targets for wise transportation is to carry out unified management and mining analysis of the huge number of the multivariate and heterogeneous dynamic transportation information, provides real-time transportation information, increase the utilization efficiency of transportation, promote transportation management and service level of travel information and provide decision support of transportation management by virtual reality as visual.

## I. INTRODUCTION

The appearance of new technologies such as internet of things and cloud computing brings opportunity for the development of wise transportation. Wise transportation is the efficient integration of intelligent sensor technology, information internet technique, communication transmission technology and data processing technique represented by car networking and cloud computing and then it is applied in the whole transportation system. It is the comprehensive transportation system which plays its role in a larger space-time scope. Compared with intelligent transportation, wise transportation not only accumulates and delivers the data but emphasizes more on the utilization and exploitation of data and focuses more on the traffic analysis, knowledge discovery, and decision-making reaction of information [29]. Some tasks in which traditions are replaced by intelligent technology require manual discrimination and resolution to reach the optimization. In addition, under the development of car networking, wise transportation emphasizes more on the interconnection of transportation information system with other information systems the furthest. Then we can see that further efficient management and deep analysis of transportation data is the key task to develop wise transportation. It requires the collection and integration of the transportation information from many fields and then the information is fully excavated to play its role in easing traffic jam, guaranteeing transportation security and ensuring the quick and environmental transportation [21].

Facing various traffic actors, cloud service platform for wise transportation provides timely comprehensive service platform which enriches transportation information [6]. The platform construction relies on the increasingly mature cloud computing technology and internet of things and solves the evils radically which are insufficient processing capacity of traditional information and platform information and the poor channel of information interaction [1]. Traffic data has several characteristics, such as large scale, diverse predictable and real-time, which falls in the range of definition of Big Data [2].

## II. SYSTEM DIVISION

The cloud service platform target for wise transportation is to carry out unified management and mining analysis of the huge number of the multivariate and heterogeneous dynamic transportation information [4], provide real-time transportation information, increase the utilization efficiency of transportation, promote transportation management and service level of travel information and provide decision support of transportation management by virtual reality as visual [5]. The concrete research goals are as follows:

1) Integrate the transportation fully, conduct processing and management about the multivariate and heterogeneous transportation data and in the meanwhile, support the interconnection between transportation information and other information systems to promote the interactive cooperation between the various functional departments and industrial departments of transportation and promote the sharing and maximum utilization of resources.

2) Realizing the storage and computation of huge amount of transportation information by utilizing cloud platform not only solves the practical problem that transportation departments cant store huge amount of transportation information but also realizes the real-time computation and analysis of the huge amount of transportation information through the strong computing ability of cloud platform and provides real-time and stable networking service for wise transportation department.

3) Conducting deep data excavation and analysis about the multi-source data makes the traffic warden and transportation participants master and understand not only the timely transportation condition, but also the variation trend of transportation condition and abnormal identification by keeping comparing and analyzing the real-time data and historical data. Through the long-term analysis, it carries out early demonstration about transportation organization and plan and provides correct, informative data support and imitation for the decision of the leadership.

The 3D visualization of city is to take inventory and display various types of object data management and resource data within the community area which is so-called virtual community [11] based on RIA technology [26]. The 2D statistical analysis visualization is overlapped with a white background on 3D virtual reality environment, since it's more intuitive and a cognitively less demanding display system, which lessens the cognitive workload of the user [22]. The proposed platform is based on WebVRGIS [19] and WebVR engine [13], which is extended to virtual city [18] [9], and is used to analysis and forecast city traffic [8] [17].

## III. KEY TECHNOLOGY OF THE PLATFORM

### A. Fault-tolerance Processing of Transportation Information

Due to the work breakdown and error of the traffic sensor and transmission equipment, its inevitable that mistakes and losses occur in transportation data. Therefore, the original data must be verified to get rid of the abnormal data and amend the incomplete data so as to guarantee the integrity and correctness of the data. The main research of fault-tolerance processing includes judging missing data, identifying abnormal data and amending incomplete data. Judging missing data is to keep scanning the dynamic data in a certain period according to the stated data collection time and format. Identifying abnormal data needs to distinguish whether the equipment fault data is the correct traffic abnormal data caused by traffic accident or abnormal weather. Amending incomplete data requires the utilization of mathematic model for analysis and prediction so as to supplement the missing data.

### B. The Integration of Multivariate and Heterogeneous Transportation Data

The collection methods of transportation data include probe vehicle, camera, microwave, and manual work. As the source data collected owns the different data structure, accuracy, position and timing, the multivariate and heterogeneous transportation data must be integrated before the transportation analysis. Along with the development of intelligent sensor technology and network communication technology of car networking, the multivariate heterogeneity of transportation data is becoming increasingly prominent and the relevant challenges of data integration are also more arduous. The main research of data integration is to take full advantage of various data by improving the accuracy of the data and lifting the network coverage range. On the one hand, conduct complementation, cross checking and superposition calculation toward temporal-spatial data through various data model and statistical model and carry out processing about multi-source data to form a more comprehensive transportation description; on the other hand, taking discrete vehicles as the basic description unit, conduct a more accurate data integration toward the collection section from the micro perspective so as to gain more accurate transportation information.

### C. The Storage and Calculation of Mass Transportation Information

As the transportation data capacity is huge, the source is diverse and the update is frequent, storing and managing the mass data and making them satisfy the highly available and reliable requirement of transportation system application is the important technical support in construction of wise transportation. For mass data storage, the database for quick and convenient inquiry and management should be established and the high-efficient transmission under the limited band of internet should be realized. At the same time, the acquisition, integration, analysis and application of mass transportation data requires the support of a high-powered computer platform. Also it needs to provide the same powerful computing ability and network service as 'super computer' through the construction of government service cloud platform.

### D. The Analysis and Prediction of Real-time Dynamic Transportation Flow

Real-time analysis and prediction of transportation flow are always the difficult points in transportation analysis. Since the change of transportation network condition is fast and complicated, along with the increase of update speed of dynamic transportation data, it has higher requirement about the traffic capacity for real-time analysis and the prediction of various transportation methods across the internet. The main research contents include analyzing the congestion degree of road network, judging the severity of traffic accident, predicting the space-time impact scope of traffic accident and the road speed of various transportation means across the internet after several minutes and hours based on the data integration.

### E. The Three-dimensional Analysis of Real-time Transportation Information

Along with the development of three-dimensional geographic information system and peoples visual requirement about the real three-dimensional scene, its the inevitable trend to establish the three-dimensional geographic information system. The main research contents include three-dimensional visual representation and inquiry of the layering of transportation network flow at different scales, three-dimensional visual representation and inquiry about various transportation accidents, the visual support of decision scheme and the analysis of model research based on space of three-dimensional data model.

## IV. THE ADVANTAGE OF PLATFORM SOLUTION

### A. High Usability and High Stability

Guaranteeing the safe and stable operation of the system by utilizing virtualization, distributed computation and cloud technology increases the safety and disaster-toleration of the system and data the furthest.

### B. High Performance and Real-time Performance

Breaking the performance bottleneck of traditional architecture based on the system architecture of cloud computing: real-timing monitoring, intelligent monitoring of transportation information, easy processing of mass data which satisfies the high-performance requirement of real-time processing of mass dynamic transportation information.

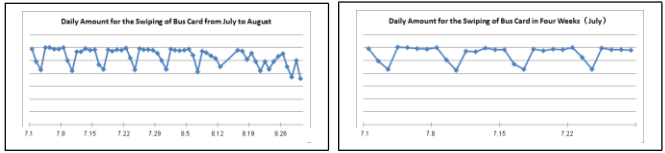

Fig. 1. Left: The statistics of the daily amount for the swiping of bus card from July to August; Right: Daily amount for the swiping of bus card in four weeks(July).

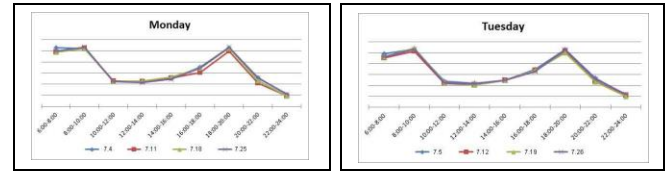

Fig. 2. Left: The statistic figure for the amount of the swiping of bus card during various periods on Monday in July; Right: The statistic chart for the amount of the swiping of bus card during various periods on Tuesday in July.

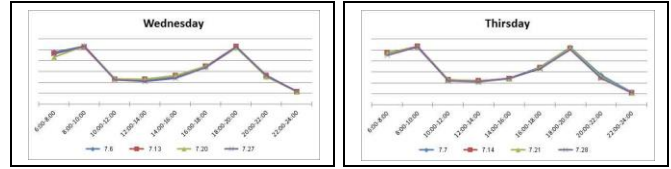

Fig. 3. Left: The statistic chart for the amount of the swiping of bus card during various periods on Wednesday in July; Right: The statistic chart for the amount of the swiping of bus card during various periods on Thursday in July.

*C. High compatibility and easy access to information*

Application load has strong adaptability, extensive compatibility, easy expansion. The compatibility includes a variety of channels for the method of issuing information such as PSTN, internet, 3G wireless which support the issue terminal of information such as PC, PDA, tablet PC.

## V. ANALYSIS OF BUS PASSENGER FLOW IN THE CITY

*A. The Analysis of Bus Service Scope*

At present, there are 47 city bus lines which pass Fukuda transportation junction. If residents choose to travel by the bus route 500 meters far from them under most circumstances, then taking the 500 meters on the two sides of the 47 bus routes as the distance for buffer zone, then the acreage of the buffer area can be calculated and it is about 400 square meters. That is the relevant nonstop bus routes to Zhuzilin can serve the residents in the area of about 400 square meters in Shenzhen.

*B. The Relevant Bus Flow Analysis of the Junction*

The bus card data of Shenzhen Tong includes ID(Card Numbers of Shenzhen tong held by the passenger), type(the vehicle types taken by the passengers). For example, '21' means swiping card for subway entry, '22' means swiping card for subway exit, '31' means swiping card for getting on the bus, DeviceID(the number of the machine for the passenger to swipe the card), strTime(the time when the passenger gets on the bus and swipe the card), Busline(the bus route taken). As theres only the storage for the time when passenger swipes the card and the bus route taken and there is no information about the getting-on station, getting-off station and getting-off time of the passengers. Therefore, the bus card data of Shenzhen Tong does not support the correct bus flow analysis related to the junction. So this research only analyzes the 47 bus routes related to Fukuda junction and the flow law in all the stations to predict the bus flow in this junction. Figure 1 shows the daily flow statistics of the bus route which is directly linked to Fukuda transportation junction. We can see that apart from the special holiday and large-scale activities, the relevant bus flow changes regularly over a seven-day period.

The following is the statistics of the passenger flow during various periods on the same day every week in four weeks from July 1 to July 28 and the periodicity is obvious.

On the workday from Monday to Friday, it presents the obvious peak phenomenon in the morning and evening and the morning peak phenomenon, which goes from 6 am to 8 am and 8 am to 10 am, lasts longer. During the peak period, the amount for the swiping of bus card is 10000 every two hours. During the same period, the amount for the swiping of bus card almost stays the same with no obvious growth trend.

## VI. THE PASSENGER FLOW ANALYSIS OF PUBLIC TRANSPORTATION TRANSFER

Public transportation transfer includes many modes such as bus-bus, subway-bus, bus-water transportation. Theoretically, its the exponential function form of n public transportation modes which regards 2 as the base number. But in a specific transportation junction, there are only a few kinds[41]. In the concrete research for Zhuzilin transportation junction, based on the construction designing structure and geographic location of Zhuzilin and combining with the characteristics of land utilization nearby and the space-time distribution of the existing passenger flow data, the several definite transfer

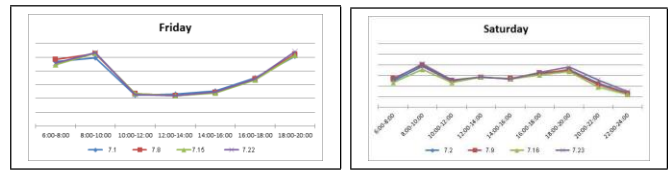

Fig. 4. Left: The statistic chart for the amount of the swiping of bus card during various periods on Friday in July; Right: The statistic chart for the amount of the swiping of bus card during various periods on Saturday in July.

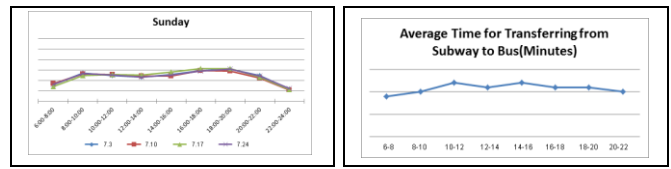

Fig. 5. Left: The statistic chart for the amount of the swiping of bus card during various periods on Saturday in July; Right: Average time for the transfer from subway to bus during various periods.

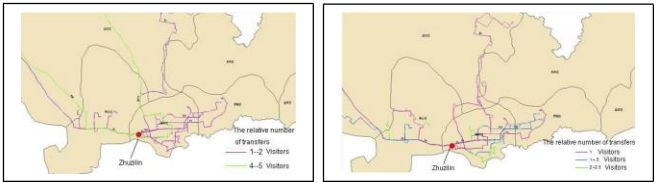

Fig. 6. Left: The main transfer bus routes after getting off the Zhuzilin subway station and relative transfer volume of the routes; Right: The bus route taken bafore transferring the subway in Zhuzilin and the relative transfer volume of various routes.

modes for key research are shown as the followings:

- transfer from a subway to a bus and the transfer station is Zhuzilin.
- transfer from a bus to a subway and the transfer station is Zhuzilin.
- transfer from a bus to a subway, the transfer station is one station of the metro, get off the subway at Zhuzilin Station.

Other transfer modes such as transferring a bus from another are not discussed in the research in this period as there is no support from 'Shenzhen Tong' data. The following contents analyze the three transfer modes in order to gain the transfer modes in Zhuzilin juncture, the inner structure of the mode and the mutual relations, etc.

The transfer mode of taking the subway to Zhuzilin and transferring to another bus is an important transfer mode. This research defines the records of the swiping of bus card within half an hour after swiping the bus card when getting out of the Zhuzilin subway station as a transfer from the subway to the bus. Figure 6 left shows the 10 bus routes location distribution in the map for transferring after getting off the subway in Zuzilin. The color of the line means the quantity of the transfer passenger flow. Relative transfer volume means setting up a standard for the number of the transfer passengers. Relative transfer volume which is the ratio of the actual number of transfer passengers for relevant buses in Zhuzilin and the standard for the number of the transfer passengers helps show the proportion structural relationship of relevant amount of bus transfer passengers. There are many kinds of traffic passenger flow forecasting models, and the common models include regression forecasting model and time series prediction model [7] [10].

Taking the bus to Zhuzilin and then taking subway in Zhulin to reach the destination is also an important transfer modes. The data sources can be the 'Shenzhen Tong' records in which passengers swipe the card after taking the bus, and then enter Zhuzilin subway station by swiping the card again within one hour. Figure 6 right is the geographic distribution of bus routes in this transfer mode. The different color of the line means the quantity of the relative transfer volume.

The third transfer mode is the secondary transfer mode that passengers firstly take a bus, then transfer to Luobao route with the destination being Zhuzilin transportation juncture. The data sources can be the 'Shenzhen Tong' records in which passengers swipe the bus card after taking the buses, swipe the car again when they take the Luobao route and get out of

TABLE I. THE PROPORTION OF THE MAIN TRANSFER BUS PASSENGER VOLUME TO ZHUZILIN FROM THE TRANSFER OF SUBWAY

| The main bus route taken firstly | Percentage |
|---|---|
| 392 | 5% |
| 43 | 2.5% |
| 66 | 1.8% |
| B728 | 1.75% |
| 338 | 1.5% |
| 392 interval line | 1.25% |
| 113 | 1.25% |
| 70 | 1.25% |
| M250 | 1.25% |
| 3 | 1.21% |

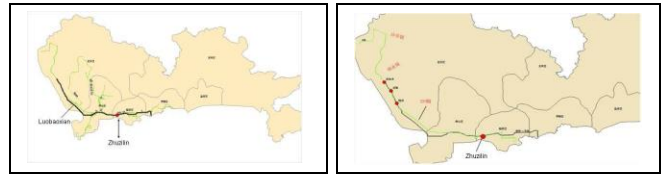

Fig. 7. Left: The 10 main bus routes to Zhuzilin by taking bus; Right: The geographic distribution diagram for No 338 bus and Luobao route.

the station in Zhuzilin after swiping the bus card within the one hour. After analysis, we get the 10 main bus routes taken and they are shown as in Table I:

The geographic distribution of the ten main routes is shown as the following. The distribution of transfer passenger flow is relatively average. It looks like that No. 338 bus parallels the Luobao route and they run about 30 kilometers in coincidence, but still in the secondary transfer mode in which passengers firstly take the bus and then transfer to the Luobao route with the destination being Zhuzilin transportation juncture, it is still the top five route. After tracking the one-card record of the passengers who adopt this transfer mode, we find that after passengers get off the bus, about 80% of them get on the subway Luobao route from three stations of East Airport, Hourui, Guwu. The three stations locate on one side of the terminal or the neighborhood in Luobao route. Then we can understand it like this: take the No.338 bus in areas in Shajin Town and Fuyongzhen Town instead of reaching Zhuzilin juncture directly. From the analysis of one-card data, it costs 40 to 50 minutes from East Airport, Hourui, Guwu to Zhuzilin. But if passengers take No. 338 bus to Zhuzilin directly, it takes 90 minutes for 30 kilometers with the bus speed being no more than 18 kilometers an hour. In order to avoid the congestion in peak period and arrive at Zhuzilin on time, passengers may get off the bus and transfer to subway Luobao route. Then we can see that the transfer of various transportation means can compensate the transportation blind spots at some degree and improve travel efficiency. The geographical distribution of No. 338 bus and Luobao router is shown as the following:

When passengers swipe the bus card after getting off the subway and transfer to the bus, the average time is 11 minutes through statistics. We can obtain the Figure 7 after further statistics as in Table I.

As shown in Figure 5 right, we can see that when the passengers swipe the bus card after getting off the subway and transfer to the bus, the average timing is 6 am to 8 am and 8 am to 10 am in the morning peak. During the two transfer periods, the average time for transferring is 9 and 19 minutes.

TABLE II. THE AVERAGE TIME FOR TRANSFERRING AND STANDARD DEVIATION DURING VARIOUS PERIODS

| Period | Average time for transferring | Standard deviation of transfer time |
| --- | --- | --- |
| 6 am-8 am | 9 | 6.380525236 |
| 8 am-10 am | 10 | 7.071766835 |
| 10 am-12 am | 12 | 7.297721547 |
| 12 am-2 pm | 11 | 6.848422948 |
| 2 pm-4 pm | 12 | 7.229369224 |
| 4 pm-6 pm | 11 | 7.075560019 |
| 6 pm-8 pm | 11 | 6.794950708 |
| 8 pm-10 pm | 10 | 6.198755368 |

From 4 pm to 6 pm and 6 pm to 8 pm in the evening peak, the average transferring time in the two periods is 11 minutes. However, the average time for transferring in non-peak period is longer than that in peak period. For example, from 10 am to 12 am and 2 pm to 4 pm, the transferring time is 12 minutes. Then we can infer that at present the transfer passenger flow in Zhuzilin transportation junction has not reached the designing saturation capacity, the transfer order is good and the transfer efficiency is high. Transfer time in off-peak period is 1 to 2 minutes longer than the transfer time in peak period. we can understand it like this: 1. During the peak period, the transfer passengers are on or off duty in fixed routes and they are familiar with the route and have clear goal. Under the situation when the transfer route is not crowed, quickening the footstep on purpose can shorten the time for swiping the card efficiently. 2. During peak period, there are more buses for the same bus route, which reduces the passengers waiting time in the bus station.

VII. FUTURE WORK

Through long-term monitoring and analysis, the long-term passenger flow forecasting are established under the condition of combing with economic development and urban planning. Our future analysis will be made on population travel behavior, taxi route, degree of influence on public bus, and road travelling speed under special weather conditions (such as rainstorm, typhoon, and heavy fog, etc.) The deeper data mining will be made, such as emergency evacuation aided decision support, monitoring and forecasting on large-scale group event, assisting crowd and vehicle evacuation under emergency. In addition to city, ocean data will be integrated into the proposed platform [23] [20], in which we will employ spatiotemporal visualization as the representative approach [28]. Some novel interaction approaches [12] [16] [15] [14] are considered to integrate in our future work. Some novel interaction approaches are considered to be integrated in our future work [12] [15] [14]. The new network data management algorithm [24] [25] , smart grid system [3] [27],

ACKNOWLEDGMENTS

The authors are thankful to the National Natural Science Fund for the Youth of China (41301439) and Electricity 863 project(SS2015AA050201), Shenzhen S&R development Fund (CXZZ20130321092415392).

REFERENCES

[1] M. Breunig and S. Zlatanova. Review: 3d geo-database research: Retrospective and future directions. *Comput. Geosci.*, 37(7):791–803, July 2011.

[2] F. Briggs. Large data - great opportunities. Presented at IDF2012, Beijing, 2012.

[3] L. Che and M. Shahidehpour. Dc microgrids: Economic operation and enhancement of resilience by hierarchical control. *IEEE Transactions Smart Grid*, 2014.

[4] E. Chow, A. Hammad, and P. Gauthier. Multi-touch screens for navigating 3d virtual environments in participatory urban planning. In *CHI '11 Extended Abstracts on Human Factors in Computing Systems*, CHI EA '11, pages 2395–2400, New York, NY, USA, 2011. ACM.

[5] C. DiSalvo and J. Vertesi. Imaging the city: Exploring the practices and technologies of representing the urban environment in hci. In *CHI '07 Extended Abstracts on Human Factors in Computing Systems*, CHI EA '07, pages 2829–2832, New York, NY, USA, 2007. ACM.

[6] A. Etches, D. Parker, S. Ince, and P. James. Utis (urban transportation information system) a geo-spatial transport database. In *Proceedings of the 8th ACM International Symposium on Advances in Geographic Information Systems*, GIS '00, pages 83–88, New York, NY, USA, 2000. ACM.

[7] H.-Z. Li, S. Guo, C.-J. Li, and J.-Q. Sun. A hybrid annual power load forecasting model based on generalized regression neural network with fruit fly optimization algorithm. *Know.-Based Syst.*, 37:378–387, Jan. 2013.

[8] X. Li, Z. Lv, J. Hu, B. Zhang, L. Yin, C. Zhong, W. Wang, and S. Feng. Traffic management and forecasting system based on 3d gis. In *15th IEEE/ACM International Symposium on Cluster, Cloud and Grid Computing (CCGrid)*, 2015.

[9] X. Li, Z. Lv, B. Zhang, W. Wang, S. Feng, and J. Hu. Webvrgis based city bigdata 3d visualization and analysis. In *IEEE Pacific Visualization Symposium (PacificVis)*, 2015.

[10] K.-P. Lin, P.-F. Pai, Y.-M. Lu, and P.-T. Chang. Revenue forecasting using a least-squares support vector regression model in a fuzzy environment. *Inf. Sci.*, 220:196–209, Jan. 2013.

[11] Z. Lu, S. U. Rehman, and G. Chen. Webvrgis: Webgis based interactive online 3d virtual community. In *Virtual Reality and Visualization (ICVRV), 2013 International Conference on*, pages 94–99. IEEE, 2013.

[12] Z. Lv. Wearable smartphone: Wearable hybrid framework for hand and foot gesture interaction on smartphone. In *2013 IEEE International Conference on Computer Vision Workshops*, pages 436–443. IEEE, 2013.

[13] Z. Lv and et.al. Webvr - - web virtual reality engine based on P2P network. *JNW*, 6(7):990–998, 2011.

[14] Z. Lv, L. Feng, S. Feng, and H. Li. Extending touch-less interaction on vision based wearable device. In *IEEE Virtual Reality Conference 2015, 23 - 27 March, Arles, France*, 2015.

[15] Z. Lv, L. Feng, H. Li, and S. Feng. Hand-free motion interaction on google glass. In *SIGGRAPH Asia 2014 Mobile Graphics and Interactive Applications*. ACM, 2014.

[16] Z. Lv, A. Halawani, S. Feng, H. Li, and S. U. Rehman. Multimodal hand and foot gesture interaction for handheld devices. *ACM Trans. Multimedia Comput. Commun. Appl.*, 11(1s):10:1–10:19, Oct. 2014.

[17] Z. Lv, X. Li, J. Hu, L. Yin, B. Zhang, and S. Feng. Virtual geographic environment based coach passenger flow forecasting. In *IEEE Computational Intelligence and Virtual Environments for Measurement Systems and Applications (CIVEMSA)*, 2015.

[18] Z. Lv, X. Li, B. Zhang, W. Wang, S. Feng, and J. Hu. Big city 3d visual analysis. In *36th Annual Conference of the European Association for Computer Graphics (Eurographics2015)*, 2015.

[19] Z. Lv, S. Rhman, and G. Chen. Webvrgis: A p2p network engine for vr data and gis analysis. In M. Lee, A. Hirose, Z.-G. Hou, and R. Kil, editors, *Neural Information Processing*, volume 8226 of *Lecture Notes in Computer Science*, pages 503–510. Springer Berlin Heidelberg, 2013.

[20] Z. Lv and T. Su. 3d seabed modeling and visualization on ubiquitous context. In *SIGGRAPH Asia 2014 Posters*, page 33. ACM, 2014.

[21] H. J. Miller and Y.-H. Wu. Gis software for measuring space-time accessibility in transportation planning and analysis. *Geoinformatica*, 4(2):141–159, June 2000.

[22] T. Porathe and J. Prison. Design of human-map system interaction. In *CHI '08 Extended Abstracts on Human Factors in Computing Systems*, CHI EA '08, pages 2859–2864, New York, NY, USA, 2008. ACM.

[23] T. Su, Z. Lv, S. Gao, X. Li, and H. Lv. 3d seabed: 3d modeling and visualization platform for the seabed. In *Multimedia and Expo Workshops (ICMEW), 2014 IEEE International Conference on*, pages 1–6. IEEE, 2014.

[24] Y. Wang, W. Jiang, and G. Agrawal. Scimate: A novel mapreduce-like framework for multiple scientific data formats. In *CCGrid2012*. IEEE, 2012.

[25] Y. Wang, A. Nandi, and G. Agrawal. Saga: Array storage as a db with support for structural aggregations. In *SSDBM '14*, New York, NY, USA, 2014. ACM.

[26] M. Zhang, Z. Lv, X. Zhang, G. Chen, and K. Zhang. Research and application of the 3d virtual community based on webvr and ria. *Computer and Information Science*, 2(1):P84, 2009.

[27] M. Zhang, Y. Sun, S. Dang, and K. Petrou. Smart grid-oriented algorithm of data retrieval and processing based on crio. In *ISEEE2014*. IEEE, 2014.

[28] C. Zhong, T. Wang, W. Zeng, and S. M. Arisona. Spatiotemporal visualisation: A survey and outlook. In *Digital Urban Modeling and Simulation*, pages 299–317. Springer, 2012.

[29] X. Zhou, D. Di, X. Yang, and D. Wu. Location optimization of urban passenger transportation terminal. In *Proceedings of the 2010 International Conference on Optoelectronics and Image Processing - Volume 01*, ICOIP '10, pages 668–671, Washington, DC, USA, 2010. IEEE Computer Society.